\documentclass[conference,compsoc]{IEEEtran}
\usepackage{usenix,epsfig,endnotes}
\usepackage{cite}
\usepackage{amsmath,amssymb,amsfonts}
\usepackage{algorithmic}
\usepackage{graphicx}
\usepackage{textcomp}
\usepackage{xcolor}
\usepackage{subcaption}
\usepackage{authblk}
\usepackage{enumerate}

\ifCLASSINFOpdf
\else
\fi
\begin{document}
%
\title{Data Augmentation for Opcode Sequence Based Malware Detection}

\author[ ]{Niall McLaughlin}
\author[ ]{Jesus Martinez del Rincon}
\affil[ ]{Centre for Secure Information Technologies  (CSIT)\\ Queen's University Belfast}
\affil[ ]{\textit {\{n.mclaughlin, j.martinez-del-rincon\}@qub.ac.uk}}

\maketitle

\begin{abstract}
In this paper we study data augmentation for opcode sequence based Android malware detection. Data augmentation has been successfully used in many areas of deep-learning to significantly improve model performance. Typically, data augmentation simulates realistic variations in data to increase the apparent diversity of the training-set. However, for opcode-based malware analysis it is not immediately clear how to apply data augmentation. Hence we first study the use of fixed transformations, then progress to adaptive methods. We propose a novel data augmentation method -- Self-Embedding Language Model Augmentation -- that uses a malware detection network's own opcode embedding layer to measure opcode similarity for adaptive augmentation. To the best of our knowledge this is the first paper to carry out a systematic study of different augmentation methods for opcode sequence based Android malware classification.
\end{abstract}


%
\IEEEpeerreviewmaketitle

\section{Introduction}

\begin{figure*}[ht!]
\centering
\includegraphics[width=0.8\textwidth, trim=1cm 6.6cm 10cm 5.6cm, clip=true]{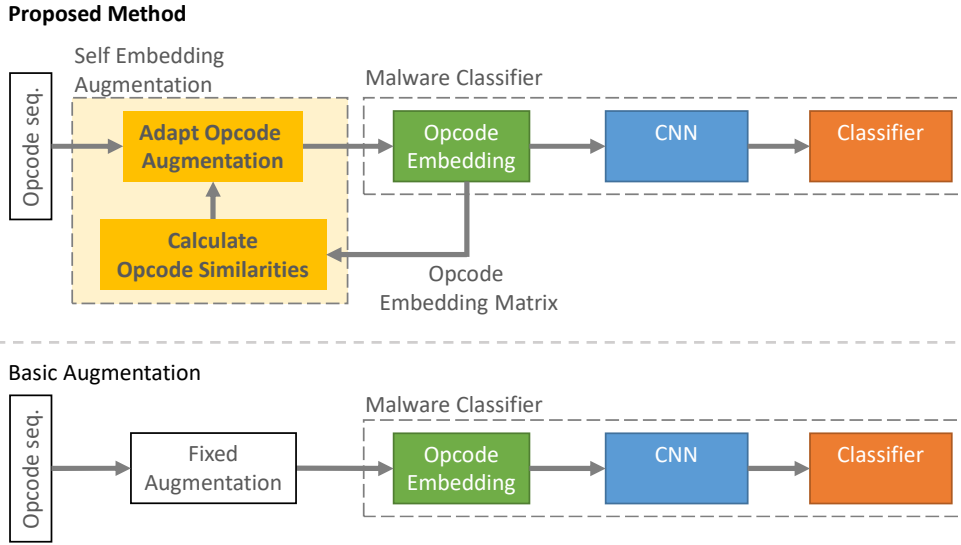}
\caption{Top: Proposed Self-Embedding Augmentation for opcode-sequence based malware classifiers. The network's own opcode embedding matrix is used to generate realistically augmented versions of opcode sequences. Augmentation parameters are updated online during network training to generate better augmentations. 
Bottom: A basic augmentation approach that uses fixed parameters.
\label{fig:proposed_system}}
\end{figure*}

Data augmentation is used to improve the generalisation of machine-learning models by artificially increasing the diversity of the training data~\cite{shorten2019survey}. Data augmentation is needed because we often have limited training data, which does not encompass the full diversity of in-the-wild data. Using augmentation we can expose the network to a larger variety of data than is present in our limited training set. The use of augmentation has gained increasing importance due to deep-learning which often requires large amounts of training data. The disadvantage of data augmentation is that the augmented samples may be highly correlated with the existing training data, so this approach is necessarily limited in the performance boost it can provide.
When applying data augmentation, the variability of samples in the original training set is artificially increased by modifying each training example using one or more transformation operations. The transformation operations are usually designed to mimic natural variations in the data. For instance, in image classification we may want an object detector to recognise a cat whether it is seen from the left or right, or rotated slightly. Hence, mirroring and/or small rotations are commonly used for vision tasks. The goal is to make the model invariant to these transformations, thus improving its robustness to similar real world variations. However not all augmentations correspond to real-world variations e.g. Mixup~\cite{zhang2017mixup}. Nevertheless, such methods have been shown empirically to improve generalisation.
For familiar types of data such as images, video, and audio, intuition can often provide guidance on the design of novel augmentation schemes. However, for more abstract data, such as opcode sequences, designing augmentation methods is more challenging. 
%
%
In this paper, we study data augmentation applied to opcode-sequence based malware classifiers~\cite{mclaughlin2017deep,millar2020dandroid}. However, our methods are general, meaning they could be applied to sequences of raw bytes e.g., Malconv~\cite{raff2018malware}, or 2D malware images~\cite{kumar2018malicious}. 
%
%
The contributions of this work are:

\begin{enumerate}[(i).]
\item We perform a systematic study of data augmentation methods applied to opcode-sequence based malware detection.
\item We introduce a novel adaptive data augmentation technique, Self-Embedding Language Model Augmentation, shown in Fig.~\ref{fig:proposed_system}. This method uses the network’s own opcode embedding layer to measure opcode similarity to apply adaptive data augmentation during training.
\end{enumerate}

\section{Related Work}

Opcode based malware detection has been extensively studied. Some of the early approaches were based on short n-grams~\cite{canfora2015mobile, kang2013android, jerome2014using} using classical machine learning algorithms. However they required extensive pre-processing and feature selection.
Recently, deep-learning architectures designed for image classification have been applied~\cite{khan2019analysis,kumar2018malicious}, where the opcode sequences is  transformed into an image for classification by existing image classification network architectures. Following from this, efficient sequence processing architectures originally proposed for natural language processing (NLP)~\cite{zhang2015character} e.g., 1D CNNs, were used. By processing sequences using local filters, the inefficiency of recursive models such as LSTMs~\cite{hochreiter1997long} is avoided. To the best of our knowledge, data augmentation has not been applied with such methods to date. This is likely due to the difficulty of designing an appropriate augmentation scheme for computer code.

Opcode-sequence based malware detection shares commonalities with NLP. Data augmentation methods in NLP include: synonym replacement, random word insertion, random word swap and random word deletion. While these techniques may give some benefit~\cite{wei-zou-2019-eda} they tend to work best on smaller datasets. In the masking technique words are replaced with a special 'blank' word~\cite{DBLP:conf/iclr/XieWLLNJN17}. This can be seen as a form of regularisation that works by adding noise to the input. A related method, replaces words using synonyms from a thesaurus~\cite{zhang2015character} to create semantically similar but different sentences. Building on the idea of substituting semantically similar words, a word embedding model such as GloVe~\cite{pennington2014glove} can be used to measure word similarity. This allows replacement of words with semantically close neighbours based on their word embeddings~\cite{jiao2020tinybert}. The NLP techniques above are not easy to apply to opcode sequences as naive substitution of opcodes may break program functionality. There are also no widely available pre-trained embeddings for opcode sequences. In the field of malware classification we are aware of only one other related work, which applies additive noise to 3-channel encoded binary file images~\cite{catak2021data} for augmentation.






\section{Method}

\subsection{Opcode Sequence Malware Classification}
\label{sec:method_cnn}

The opcode sequence of a given program can be recovered either via disassembly or by dynamic analysis. In this work we focus on static analysis by disassembly, although our methods are general enough to be applied to dynamic analysis. We use a CNN based malware classifier~\cite{mclaughlin2017deep}. The complete network consists of an opcode embedding layer, a single convolutional layer, a max pooling layer, a linear classification layer, outputting the probability a sample is malicious. This style of architecture is often referred to as \emph{Malconv} and has been successfully validated~\cite{mclaughlin2017deep,millar2020dandroid,millar2021multi}.

\subsection{Data Augmentation of Opcode Sequences}

We use the term data augmentation to describe the technique where the raw inputs to a network are modified during training. The inputs are usually modified to reflect variations the network should be invariant to during testing~\cite{shorten2019survey}. In a sense, data augmentation artificially increases the size of the training-set and is commonly applied to improve the network's generalisation performance by showing it a greater variety of examples than those present in the training-set.

Data augmentation can be applied either offline or online~\cite{shorten2019survey, chilimbi2014project}. In online augmentation every training example is randomly modified in a different way at every training epoch, thus increasing the variety of training data. Offline augmentation is typically applied when the augmentation is too computationally costly to be used during the network's training loop~\cite{chilimbi2014project}. For text/sequence based inputs, such as opcodes sequences, the computational costs of online augmentation are negligible, therefore in this work we always use online augmentation.

Typically, the strength of augmentation must be carefully selected so that the network converges and generalises well on real data. There are several ways in which the 'strength' of data augmentation can be varied. Firstly,  the ratio of augmented data to un-augmented data presented to the network during training can be varied. We use the parameter $\beta$ to describe the probability the network will be presented with an augmented sample at any given epoch. Varying $\beta$ varies the weighting of the learned features between augmented and un-agumented data.

Secondly, we can directly vary the strength of the augmentation methods themselves. This parameter will depend on the specifics of each augmentation technique. For example, we can vary the percentage of opcodes randomly replaced with zeros (See Section~\ref{sec:input_dropout}). We use parameter $\alpha$ to denote the strength of the augmentation. 

Finally, we can vary whether to use data augmentation during training only, or whether to use test-time augmentation~\cite{shanmugam2020and}. In test-time augmentation, multiple augmented samples are passed through the network at inference time and the final prediction based on aggregating the multiple predictions. Test-time augmentation can improve performance in some cases~\cite{krizhevsky2012imagenet,shanmugam2020and}. In this work we only use data augmentation during training, which simplifies our analysis. Our reported performance figures therefore represent a lower bound and may be further improved by the use of test-time augmentation.

\subsection{Basic Opcode Sequence Augmentation Methods}
\label{sec:simple_opcode_aug}
We will now introduce several basic methods of opcode sequence based data augmentation. These methods do not adapt to the context of a particular sequence, and are based on either heuristic principles, or existing approaches that have successfully been applied to text-based data augmentation in NLP~\cite{wei-zou-2019-eda,DBLP:conf/iclr/XieWLLNJN17,jiao2020tinybert}.

\paragraph{\textbf{Input Dropout}}
\label{sec:input_dropout}
%
A random fraction of opcodes in each program are replaced with zeros. Opcodes for replacement are selected uniformly at random from the whole opcode sequence. This can be seen as a form of dropout~\cite{srivastava2014dropout} applied directly to the input. It is similar to methods that add noise to the input sequence to improve generalisation~\cite{DBLP:conf/iclr/XieWLLNJN17}. The strength of augmentation can be varied by varying the hyperparameter, $\alpha$, which specifies the fraction of opcodes to randomly replace. 

Note that in the case of opcode sequences, this operation is not the same as replacing opcodes with the {\fontfamily{qcr}\selectfont nop} (no operation) instruction. We reserve as a special ‘blank’ character used only for augmentation with zeros. This 'blank' character does not appear anywhere in the original opcode sequences, so it has no semantic significance in terms of the programmatic code, and instead simply represents missing information.

\paragraph{\textbf{Random Replacement}}
A fraction of opcodes selected uniformly at random is replaced with different opcodes selected uniformly at random from the instruction set. The hyperparameter $\alpha$ specifies the fraction of opcodes to randomly replace, hence it represents the augmentation strength. This augmentation method increase the noise present in the opcode sequence and may force the network to learn to ignore outlier opcodes in the opcode sequence or to ignore short unrealistic opcode sequences.

\paragraph{\textbf{Similar Instructions}}
\label{subsec:similar_instructions}

A fraction of opcodes in each program are randomly replaced with an instruction selected from a hand-designed list of semantically similar instructions. Opcodes to be replaced are selected uniformly at random from the opcode sequence.

For every instruction in the instruction-set, a list of similar instructions was created using the names of instructions in the Android virtual machine instruction-set. For instance, instructions such as   {\fontfamily{qcr}\selectfont move, move/from-16, move-16} etc. can be regarded as semantically similar. For each instruction, a table of similar instructions was created based on the instruction prefixes e.g.: {\fontfamily{qcr}\selectfont move, const, goto, cmp, if, get, put, cast} etc.
Instructions without obvious semantic neighbours are augmented using the random replacement method. This augmentation technique is intended to create novel, but semantically similar, malware samples to ensure better generalisation of the classifier. The hyperparameter $\alpha$ specifies the fraction of opcodes in a given program to replace, hence it varies the augmentation strength.

\paragraph{\textbf{Correlated Input Dropout}}

One possible issue with the above data augmentation methods such as input dropout, random replacement and similar instructions, is they produce uncorrelated variations in the input sequence. This is because opcodes to be replaced are selected uniformly at random. This method of augmentation may not create enough variation to pose a challenge to the classifier, and therefore may not significantly improve performance. To increase the strength and difficulty of the augmentation we instead examine the use of correlated augmentations. 

Rather than replacing individual opcodes uniformly at random we instead examine correlated opcode replacement. We select one or more instructions from the instruction-set, then replacing all instances of these instructions in the opcode sequence with the special reserved 'blank' character. For example, if we select the move instruction, all instances of the move instruction in the opcode-sequence would be replaced with 'blank'. The same process can be repeated for additional instructions to increase augmentation strength.

This form of augmentation has a significant effect on the program semantics and may force the network to learn more robust features, as it cannot rely on any particular instruction in isolation to perform classification. It must instead learn more robust opcode patterns. This method differs from Input Dropout as it forces the network to learn with significant amounts of correlated noise. The parameter $\alpha$ controls the number of the instructions to replace as a faction of the total size of the processor's instruction-set, hence varies augmentation strength.

\subsection{Adaptive Opcode Sequence Augmentation Methods}
\label{sec:adaptive_opcode_aug}

In this section we introduce several adaptive augmentation methods. These methods use knowledge of the semantic similarity of opcodes to augment samples in a way that may be more realistic than the basic basic augmentation methods introduced in Section~\ref{sec:simple_opcode_aug}.


\subsubsection{Language model}

First introduced for natural language processing, language models learn the semantic relationships between words in natural language text. A language model can be used to measure the semantic similarity of different words. Language models can be trained to predict a missing character/word in a sequence of text conditioned on the neighbouring characters/words. Alternatively,they can be trained to predict the next character/word in a sentence based on the preceding words~\cite{Bengio:2008}. We adapt this idea for opcode sequence based malware analysis. Given a large number of opcode sequences, we train an off-the-shelf language model to predict a missing opcode given its neighbouring opcodes. The resulting model captures the semantic similarity of different opcodes. We use the Continuous Bag of Words (CBOW) word2vec algorithm~\cite{mikolov2013efficient} as our language model. Given a word2vec model trained on many opcode sequences, its embedding matrix contains information on the semantic similarity of different opcodes~\cite{jiao2020tinybert}. Opcodes with similar embedding vectors tend to be semantically similar. We can then perform augmentation by replacing random opcodes in a given sequence with their semantic equivalents.

%
%

Concretely,  word2vec model is trained offline using all the opcode sequences in the training-set. After training, we extract the word2vec opcode embedding matrix. For every opcode we create a ranked list of its most similar opcodes, based on the Euclidean distance between embedding vectors. For every opcode we record its top-10 most similar opcodes. To perform augmentation of a given input sequence, opcodes are selected uniformly at random from the sequence. Each opcode is then replaced with an opcode randomly selected from that opcode's top-10 list of semantically similar neighbours. The hyperparameter $\alpha$ specifies the fraction of opcodes in the original sequence to replace. 

%
%


\subsubsection{Self-Embedding Language Model}

As part of training a \emph{Malconv} like CNN to perform malware detection we are also learning an opcode embedding matrix, as described in Section~\ref{sec:method_cnn}. The opcode embedding matrix learns the semantic relationships between opcodes. We note that the opcode embeddings learned for malware classification may be different from those learned by a language model. Malware detection and language modelling are fundamentally different tasks and may require the network to extract different information from the opcode sequence, hence each opcodes may have different semantics in each task.

We therefore hypothesise that the embedding matrix from a network trained on malware classification may be more useful for generating realistically augmented inputs for training a malware classifier. This poses a chicken-and-egg problem where we require a trained malware classifier in order to train a new one. We resolve this problem by using the embedding matrix from the network currently being trained. We use this embedding matrix to measure the semantic similarity of opcodes and hence generate realistically augmented training samples. As the network trains, the embedding matrix used to generate augmented samples is updated, hence the augmented samples become more realistic and challenging as training progresses. To allow the network time to adapt to the increasingly challenging augmented samples we use a lagging version of the embedding matrix, updated once per training epoch.

At the start of every training epoch, a copy of the malware detection network's opcode embedding matrix is made. %
Given this opcode embedding matrix, for every opcode a list of semantically similar opcodes is constructed based on the Euclidean distance between embedding vectors. For every opcode, the top-10 most semantically similar opcodes i.e., those with smallest Euclidean distances, are recorded to produce an opcode similarity table. 
To perform augmentation of a given opcode sequence during training, a fraction of the sequence's opcodes are selected uniformly at random for replacement. The hyperparameter $\alpha$ specifies the fraction of opcodes to randomly replace. Each selected opcode is then replaced with a opcodes, selected uniformly at random, from the original opcodes lists of semantically similar neighbours. 
The complete augmentation process is illustrated in Fig.~\ref{fig:proposed_system}.

We note that when training first begins, the embedding matrix is randomly initialised, so there is no semantic information yet encoded in its weights. Hence this method essentially performs random opcode replacement during early training epochs, and gradually begins to incorporate semantic information as training progresses. This simulates a curriculum based approach and means that the strength of the augmentation gradually increases during training.

\section{Experiments}
\label{sec:experiments}


To test the effectiveness of our augmentation algorithms we performed experiments using several different malware datasets and with different network hyper-parameters. 
For all experiments we use the 1D CNN network design from~\cite{mclaughlin2017deep}, which is similar to that of~\cite{raff2018malware}. The following hyper-parameters were used: Embedding Layer: $8$ dimensions, Convolutional filters: $64$, of length $8$, Max-Pooling layer, followed by a single linear layer with a $1$ dimensional output. Binary cross entropy loss was used. Adam optimiser~\cite{kingma2014adam} with batch size $48$, and learning rate of 1e-3 for $120$ epochs. During training and testing opcode sequences were truncated to $128,000$ opcodes due to GPU memory limitations. For all experiments, all code and hyper-parameters remained constant across datasets and models, with only the data augmentation method varying. Results are reported using f1-score.

Two datasets were used for the experiments: The Small Dataset, which consists of malware from the Android Malware Genome project. This dataset contains 2123 applications - 863 benign and 1260 malware samples from 49 malware families. The Large Dataset was provided by McAfee Labs (Intel Security) and consists of malware from the vendor's internal dataset. This dataset contains roughly 10,000 malware and a further 10,000 clean applications collected from the Google Play Store. Both datasets are therefore quite evenly balanced in terms of malware and clean programs.

Both datasets were thoroughly checked and cleaned to ensure no duplicate programs that contaminated the test/training splits. For experimentation purposes, each dataset was split into 5 non-overlapping folds and cross validation was preformed during evaluation. In effect, during each round of cross validation, 80\% of the dataset was used for training and 20\% for testing. For each experiment we report the average performance across the 5 cross-validation folds in terms of f1-score.

In our experiments we vary the $\alpha$ parameter, which controls the augmentation strength, and report how this affects the classification f1-score. For all experiments we hold the parameter $\beta$, which controls the probability of applying augmentation to a given sample, constant at $\beta=0.5$ i.e., half the samples presented during training are augmented and half are unchanged. While it would be be possible to vary both parameters for all experiments, we did not explore this possibility due to the excessive computational costs.


For the language model augmentation method we use the Gensim implementation~\cite{rehurek_lrec} of word2vec~\cite{mikolov2013efficient}. We use the default parameters i.e., a continuous bag-of-words (CBOW) model trained using negative sampling. We use word2vec embedding vectors of dimensionality $8$ in order to match the embedding vector size of the malware detector network. The context window size is $5$ and the model was trained for $5$ epochs. For each experiment the word2vec model was pre-trained offline using the same training data as the main malware classifier. The embedding matrix was then extracted and used to build an opcode similarity table for generating augmentations.

\subsection{Baseline Performance}

Baseline performance of the 1D CNN opcode-based malware detector network, with no data augmentation, was measured on both the Small and Large datasets. 
%
%
The average f1-score from 5-fold CV was recorded for each dataset. For all remaining experiments, all network hyper-parameters, training and testing settings and code, except that related to data augmentation, remained identical. Baseline results are shown in Table~\ref{table:simple_aug_results}.

\subsection{Basic Augmentation Methods}

We compare the basic augmentation schemes with baseline no augmentation results. The basic augmentation methods make predefined changes to the opcode sequence without regard to context. They include: 'Input Dropout', 'Random Replacement', 'Similar Instructions' and 'Correlated Input Dropout' (See Section~\ref{sec:simple_opcode_aug}). The augmentation strength, $\alpha$, was systematically varied. Note that the $\alpha$ parameter controls different aspects of the strength of each augmentation method, so we cannot directly compare $\alpha$ values across different augmentation methods. 

The results of these experiments are shown in Table~\ref{table:simple_aug_results}.
Across both datasets, several augmentation methods produce consistent improvements over the baseline for a range of $\alpha$ values. In particular, Input Dropout consistently performs well with peak performance occurring at $\alpha = 0.2$ on both datasets. Several other methods, Random Replacement and Similar Instructions, outperform Input Dropout on the Small Dataset. However this performance boost is not repeated on the Large Dataset. Correlated Input Dropout performs similarly across both datasets, although its overall performance is slightly worse than Input Dropout in terms of relative improvement compared to the baseline across both datasets. The results overall suggest that these basic augmentation methods have the potential help to improve malware classification performance.


\subsection{Language Model Augmentation}



This approach generates augmented opcode sequences while aiming to preserve functionality and semantics. The gensim~\cite{rehurek_lrec} implementation of CBOW word2vec~\cite{mikolov2013efficient} with 8-dimensional opcode embedding vectors was used to calculate opcode similarity. The language model was trained offline and its embedding matrix extracted to calculate opcode similarities.

The results of the experiments are shown in Table~\ref{table:simple_aug_results}. Performance on the Small Dataset is comparable with the Self-Embedding (SE) Language Model (see Section~\ref{sec:self-embed}) method however performance on the larger dataset is similar to the basic augmentation methods.
We hypothesise that the relatively poor performance on the Large Dataset, compared to the SE Language model, may be caused by the fact that this language model is trained on a different task from malware analysis i.e., the language model is trained to predict an opcode from its context. The relationships between opcodes learned by the language model may be different those that are important for malware classification. However the overall performance of this method is better than or competitive with any of the other augmentation methods. Improving its performance may therefore present an avenue for future research.


\subsection{Self-Embedding Language Model Augmentation}
\label{sec:self-embed}


In Self-Embedding language model augmentation, the semantic similarity between opcodes for data augmentation is measured using the constantly updated embedding matrix from the malware classifier currently being trained. In these experiments the network's embedding matrix was sampled at the beginning of every epoch. Hence the table of opcode similarity is updated one per epoch.

We compare performance against the baseline system with no augmentation, and against all the other augmentation methods. Results are reported in Table~\ref{table:simple_aug_results}.
For both the Large and Small datasets we can see that augmentation by self-embedding language model has a positive effect on classification performance. Across both datasets the effectiveness of augmentation varies as the $\alpha$ value is varied. In both cases maximum performance occurs when $\alpha$ is around 0.2. Performance drops off at both higher and lower $\alpha$ values. Compared to the other augmentation methods we can see that this method has the highest performance across both datasets. When the optimal $\alpha$ value is selected, the f1-score increases by almost $1\%$ which is the largest improvement seen across all the augmentation methods. We can also see that this method consistently outperforms the baseline, showing that the value of the $\alpha$ parameter is not critical. We propose this method should be used whenever training opcode-sequence based malware classifiers that make use of an embedding matrix.

\subsection{Augmentation Performance in Context}

In this section we provide context for the performance gains achievable using data augmentation. Another way to improve the network's performance is to increase its size until just before over-fitting occurs. This comes at the cost of increased training and inference time, however it provides an upper-bound for the potential performance of the model architecture on the dataset. We therefore compare the performance of a constant sized network trained using data augmentation, with that of increasingly larger networks trained without data augmentation.



\begin{figure}[h]
     \centering
     \begin{subfigure}[b]{\columnwidth}
         \centering
         \includegraphics[width=\columnwidth,trim=2.4cm 4.1cm 2.1cm 2.2cm,clip=true]{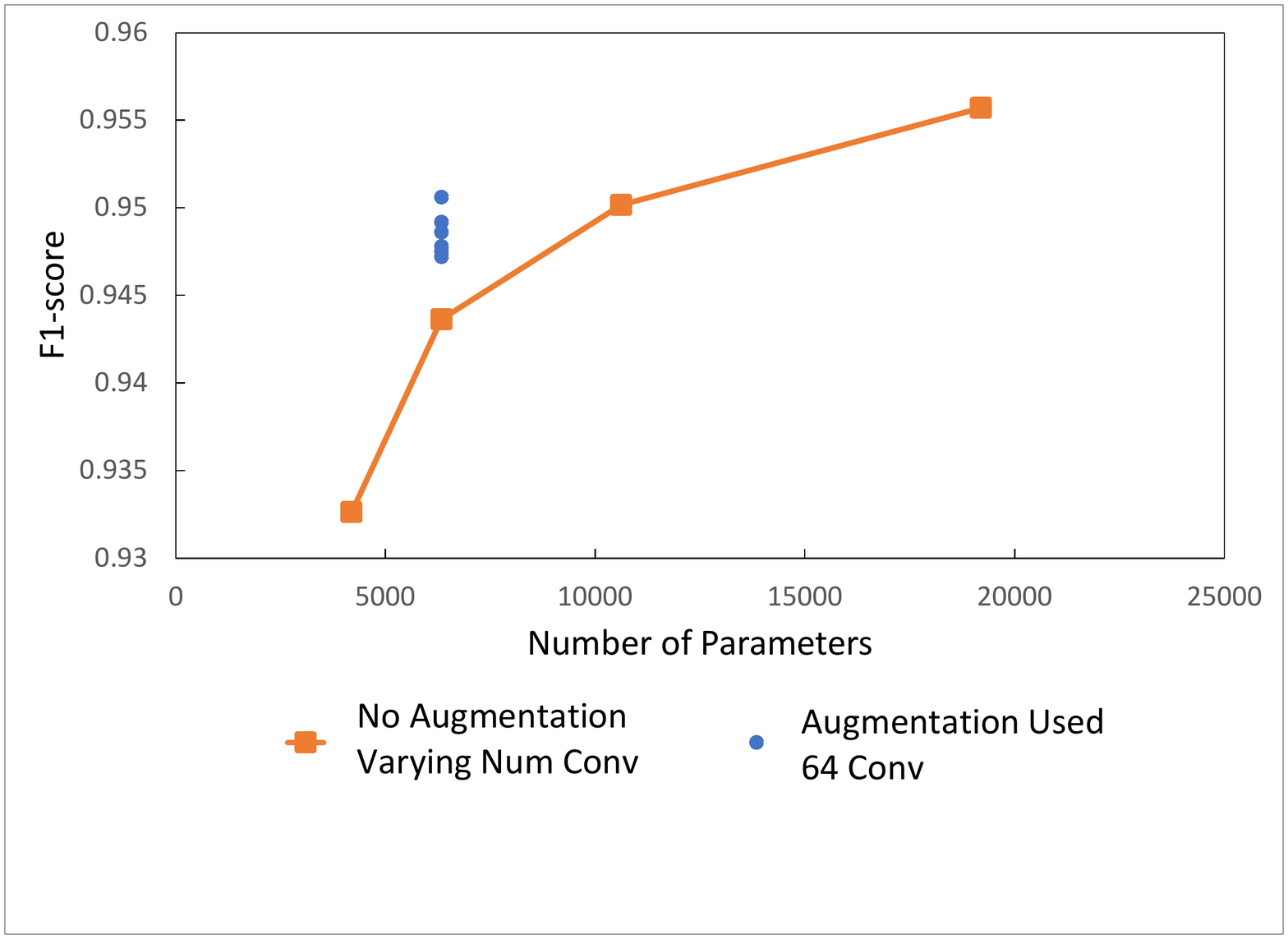}
         \caption{Comparing parameter count, and classification performance with/without data augmentation. Orange line - Networks with 32, 64, 128 and 256 convolutional filters trained without data augmentation. Blue Points - Networks with 64 convolutional filters trained using data augmentation. Data augmentation enables higher performance with smaller models 
         (Results shown for the Large Dataset).}
         \label{fig:self_embed_large_ds}
     \end{subfigure}
     \hfill
     \begin{subfigure}[b]{\columnwidth}
         \centering
         \includegraphics[width=8cm,trim=2.4cm 2.2cm 2.5cm 1cm,clip=true]{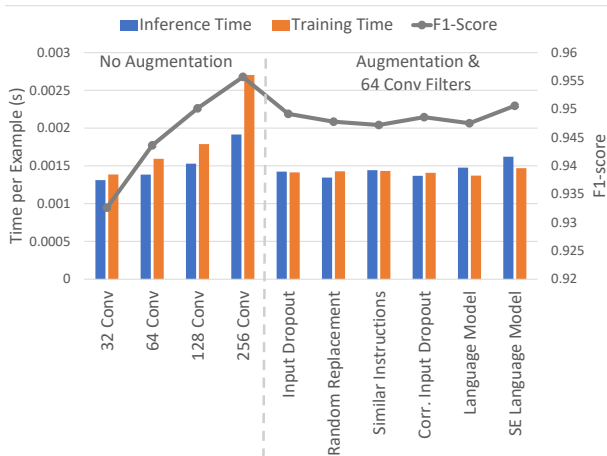}
         \caption{A 64 conv. filter network trained using data augmentation compared to networks with 32, 64, 128 and 256 conv. filters trained without data augmentation.  Inference and training time and f1-score per-example are shown. Data augmentation enables a smaller model to perform as well as a larger model with the added benefit of lower training and inference time. (Results shown for the Large Dataset).
         }
         \label{fig:self_embed_small_ds}
     \end{subfigure}
     \hfill
        \caption{Data augmentation performance in context.\label{fig:context_both}}
\end{figure}

On the Large Dataset We train networks with between 32 and 256 convolutional filters without using data augmentation. We compare these networks against a 64 convolutional filter network trained using all proposed methods of data augmentation with their optimal hyperparameters. The results are shown in Fig.~\ref{fig:context_both}. We do two types of comparison: firstly, comparing classification performance as model size is varied, and secondly, comparing classification performance versus training and inference time.

Fig.~\ref{fig:context_both} (a) shows that for models trained without data augmentation it is necessary to significantly increase the parameter count to increase performance. In contrast, by using data augmentation a networks with 64 convolutional filters can approach or exceeds the performance of a network with 128 filters but no augmentation. The self-embedding language model augmentation method even allows the smaller network to slightly exceed the performance of the 128 filters network trained without augmentation. 

Fig.~\ref{fig:context_both} (b) shows that increasing network size significantly increases both training and inference time. However for a constant network size data augmentation causes little change in training or inference time. Augmentation enables the performance the 64 convolutional filter network to slightly exceed the performance of the larger 128 filter network trained without augmentation. In addition, the smaller network achieves this with much lower training and inference time.

These experiments show that our proposed augmentation methods can significantly improve malware classification performance, without altering the memory or computation time needed. They allow a smaller network trained using augmentation to exceed the performance of a network with no augmentation.

\subsection{Combinations of Augmentations}

\begin{figure}[h]
\centering
\includegraphics[width=\columnwidth,trim=2cm 3.2cm 4.4cm 3.2cm,clip=true]{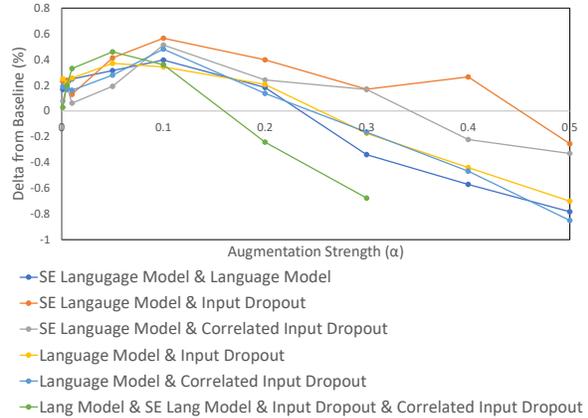}
\caption{Combinations of data augmentation methods on the Large Dataset. Each line shows the \% difference in malware classification f1-score from the baseline i.e., the network trained without data augmentation. Results are shown over a range of augmentation strengths ($\alpha$).
\label{fig:combined_aug_methods}}
\end{figure}

In this section we study the performance of combinations of augmentation methods. Using different augmentation methods together introduces more variability into the training data, which may improve generalisation. Note that we did not modify the augmentation methods to prevent interference or provide knowledge of previous augmentations applied. As above, a network with 64 convolutional filters was trained using various combined augmentations. In Fig.~\ref{fig:combined_aug_methods} we report results over a range of augmentation strengths. Note that due to the computational costs of these experiments, only a small number of combinations were tried. We selected combinations of augmentations based on those that performed well individually. These experiments were performed on the Large Dataset. 

Fig~\ref{fig:combined_aug_methods} shows that the combination of Self-Embedding Language Model and Input Dropout performed best out of the combinations tried. However this combination does not quite outperform the self-embedding language model used alone. We also find that when combinations of methods are used, the optimal augmentation strength is lower. Further research is needed to find the optimal combination of augmentation methods that complement each other.









\section{Conclusion}

In this paper we present a study of data augmentation techniques applied to opcode sequence based malware classification. Of the basic augmentation methods, the simplest method, Input Dropout, gives a consistent performance boost across both datasets. However, the adaptive method, Self Embedding Language Model, can give a larger improvement in performance compared to the basic methods. It reduced the error rate by 18\% on the small dataset and 12\% on the large dataset. Overall our work shows the potential of data augmentation for opcode sequence based malware classification. However, that further study will be required to develop more realistic augmentations and to understand the combination of basic augmentations needed to give a consistent performance improvement.

\begin{table*}[h]
\centering
\setlength\tabcolsep{3pt}
\renewcommand{\arraystretch}{1.2}
\begin{tabular}{l|cc|cc|cc|cc|ll|ll}
Method   & \multicolumn{2}{c|}{Input Dropout}     & \multicolumn{2}{c|}{Random}         & \multicolumn{2}{c|}{Similar}        & \multicolumn{2}{c|}{Correlated}     & \multicolumn{2}{c|}{Lang. Model}                       & \multicolumn{2}{c}{Self-Embed}                       \\ \hline
Dataset  & Small            & Large            & Small            & Large            & Small            & Large            & Small            & Large            & \multicolumn{1}{c}{Small} & \multicolumn{1}{c|}{Large} & \multicolumn{1}{c}{Small} & \multicolumn{1}{c}{Large} \\ \hline
Baseline & 0.94939          & 0.94363          & 0.94939          & 0.94363          & 0.94939          & 0.94363          & 0.94939          & 0.94363          & 0.94939                   & 0.94363                    & 0.94939                   & 0.94363                   \\ \hline
$\alpha$ 0.05     & 0.94877          & \textbf{0.94537} & \textbf{0.95688} & \textbf{0.94780} & \textbf{0.95549} & \textbf{0.94721} & \textbf{0.95188} & \textbf{0.94794} & \textbf{0.95835}          & \textbf{0.94866}           & \textbf{0.9502}           & \textbf{0.94666}          \\
$\alpha$ 0.1      & \textbf{0.95176} & \textbf{0.94689} & \textbf{0.94979} & \textbf{0.94595} & \textbf{0.95145} & \textbf{0.94721} & \textbf{0.9529}  & \textbf{0.94799} & \textbf{0.94990}          & \textbf{0.94829}           & \textbf{0.95118}          & \textbf{0.94926}          \\
$\alpha$ 0.2      & \textbf{0.95539} & \textbf{0.94918} & \textbf{0.95374} & \textbf{0.94647} & \textbf{0.95063} & \textbf{0.94604} & \textbf{0.95406} & \textbf{0.94860} & 0.94129                   & \textbf{0.94731}           & \textbf{0.95881}          & \textbf{0.95061}          \\
$\alpha$ 0.3      & \textbf{0.94994} & \textbf{0.94739} & 0.94049          & 0.94286          & 0.94306          & 0.94350          & \textbf{0.95220} & \textbf{0.94650} & 0.94346                   & \textbf{0.94434}           & \textbf{0.9576}           & \textbf{0.94873}          \\
$\alpha$ 0.4      & 0.94697          & \textbf{0.94681} & 0.94155          & 0.93952          & 0.94784          & 0.94353          & \textbf{0.95100} & 0.94353          & 0.93547                   & 0.94085                    & \textbf{0.95032}          & \textbf{0.94775}          \\
$\alpha$ 0.5      & 0.94487          & 0.94274          & 0.93640          & 0.93551          & 0.94018          & 0.94045          & 0.94854          & 0.94241          & 0.93510                   & 0.93908                    & \textbf{0.95267}          & \textbf{0.94827}          \\ \hline
Max      & 0.95539          & 0.94918          & 0.95688          & 0.94780          & 0.95549          & 0.94721          & 0.95406          & 0.94860          & 0.95835                   & 0.94866                    & 0.95881                   & 0.95061                   \\
Delta    & 0.00600          & 0.00555          & 0.00749          & 0.00417          & 0.00610          & 0.00358          & 0.00467          & 0.00497          & 0.00896                   & 0.00503                    & 0.00942                   & 0.00698                  
\end{tabular}
\vspace*{5mm}
\caption{Comparison of all augmentation methods as the augmentation strength ($\alpha$) is varied between 0.05 and 0.5. Results are reported as average f1-score from 5-fold cross-validation across Large and Small Datasets. Entries in \textbf{bold text} denote improved performance compared to the baseline. Max - The maximum absolute f1-score. Delta - The maximum absolute improvement over all $\alpha$ values, compared to the baseline. \label{table:simple_aug_results}}
\end{table*}



\bibliographystyle{IEEEtran}
%



\bibliography{references}

\begin{thebibliography}{10}
\providecommand{\url}[1]{#1}
\csname url@samestyle\endcsname
\providecommand{\newblock}{\relax}
\providecommand{\bibinfo}[2]{#2}
\providecommand{\BIBentrySTDinterwordspacing}{\spaceskip=0pt\relax}
\providecommand{\BIBentryALTinterwordstretchfactor}{4}
\providecommand{\BIBentryALTinterwordspacing}{\spaceskip=\fontdimen2\font plus
\BIBentryALTinterwordstretchfactor\fontdimen3\font minus
  \fontdimen4\font\relax}
\providecommand{\BIBforeignlanguage}[2]{{%
\expandafter\ifx\csname l@#1\endcsname\relax
\typeout{** WARNING: IEEEtran.bst: No hyphenation pattern has been}%
\typeout{** loaded for the language `#1'. Using the pattern for}%
\typeout{** the default language instead.}%
\else
\language=\csname l@#1\endcsname
\fi
#2}}
\providecommand{\BIBdecl}{\relax}
\BIBdecl

\bibitem{shorten2019survey}
C.~Shorten and T.~M. Khoshgoftaar, ``A survey on image data augmentation for
  deep learning,'' \emph{Journal of Big Data}, vol.~6, no.~1, pp. 1--48, 2019.

\bibitem{zhang2017mixup}
H.~Zhang, M.~Cisse, Y.~N. Dauphin, and D.~Lopez-Paz, ``mixup: Beyond empirical
  risk minimization,'' \emph{arXiv preprint arXiv:1710.09412}, 2017.

\bibitem{mclaughlin2017deep}
N.~McLaughlin, J.~Martinez~del Rincon, B.~Kang, S.~Yerima, P.~Miller, S.~Sezer,
  Y.~Safaei, E.~Trickel, Z.~Zhao, A.~Doup{\'e} \emph{et~al.}, ``Deep android
  malware detection,'' in \emph{CODASPY}, 2017, pp. 301--308.

\bibitem{millar2020dandroid}
S.~Millar, N.~McLaughlin, J.~Martinez~del Rincon, P.~Miller, and Z.~Zhao,
  ``Dandroid: A multi-view discriminative adversarial network for obfuscated
  android malware detection,'' in \emph{CODASPY}, 2020, pp. 353--364.

\bibitem{raff2018malware}
E.~Raff, J.~Barker, J.~Sylvester, R.~Brandon, B.~Catanzaro, and C.~K. Nicholas,
  ``Malware detection by eating a whole exe,'' in \emph{Workshops at the
  Thirty-Second AAAI Conference on Artificial Intelligence}, 2018.

\bibitem{kumar2018malicious}
R.~Kumar, Z.~Xiaosong, R.~U. Khan, I.~Ahad, and J.~Kumar, ``Malicious code
  detection based on image processing using deep learning,'' in
  \emph{Proceedings of the 2018 International Conference on Computing and
  Artificial Intelligence}, 2018, pp. 81--85.

\bibitem{canfora2015mobile}
G.~Canfora, F.~Mercaldo, and C.~A. Visaggio, ``Mobile malware detection using
  op-code frequency histograms,'' in \emph{2015 12th International Joint
  Conference on e-Business and Telecommunications (ICETE)}, vol.~4.\hskip 1em
  plus 0.5em minus 0.4em\relax IEEE, 2015, pp. 27--38.

\bibitem{kang2013android}
B.~Kang, B.~Kang, J.~Kim, and E.~G. Im, ``Android malware classification
  method: Dalvik bytecode frequency analysis,'' in \emph{Proceedings of the
  2013 research in adaptive and convergent systems}, 2013, pp. 349--350.

\bibitem{jerome2014using}
Q.~Jerome, K.~Allix, R.~State, and T.~Engel, ``Using opcode-sequences to detect
  malicious android applications,'' in \emph{ICC}.\hskip 1em plus 0.5em minus
  0.4em\relax IEEE, 2014, pp. 914--919.

\bibitem{khan2019analysis}
R.~U. Khan, X.~Zhang, and R.~Kumar, ``Analysis of resnet and googlenet models
  for malware detection,'' \emph{Journal of Computer Virology and Hacking
  Techniques}, vol.~15, no.~1, pp. 29--37, 2019.

\bibitem{zhang2015character}
X.~Zhang, J.~Zhao, and Y.~LeCun, ``Character-level convolutional networks for
  text classification,'' in \emph{Advances in Neural Information Processing
  Systems}, 2015, pp. 649--657.

\bibitem{hochreiter1997long}
S.~Hochreiter and J.~Schmidhuber, ``Long short-term memory,'' \emph{Neural
  computation}, vol.~9, no.~8, pp. 1735--1780, 1997.

\bibitem{wei-zou-2019-eda}
J.~Wei and K.~Zou, ``{EDA}: Easy data augmentation techniques for boosting
  performance on text classification tasks,'' in \emph{(EMNLP-IJCNLP)}.\hskip
  1em plus 0.5em minus 0.4em\relax Association for Computational Linguistics,
  2019.

\bibitem{DBLP:conf/iclr/XieWLLNJN17}
\BIBentryALTinterwordspacing
Z.~Xie, S.~I. Wang, J.~Li, D.~L{\'{e}}vy, A.~Nie, D.~Jurafsky, and A.~Y. Ng,
  ``Data noising as smoothing in neural network language models,'' in
  \emph{ICLR}.\hskip 1em plus 0.5em minus 0.4em\relax OpenReview.net, 2017.
  [Online]. Available: \url{https://openreview.net/forum?id=H1VyHY9gg}
\BIBentrySTDinterwordspacing

\bibitem{pennington2014glove}
J.~Pennington, R.~Socher, and C.~D. Manning, ``Glove: Global vectors for word
  representation,'' in \emph{Empirical Methods in Natural Language Processing
  (EMNLP)}, 2014, pp. 1532--1543.

\bibitem{jiao2020tinybert}
X.~Jiao, Y.~Yin, L.~Shang, X.~Jiang, X.~Chen, L.~Li, F.~Wang, and Q.~Liu,
  ``Tinybert: Distilling bert for natural language understanding,'' in
  \emph{Proceedings of the 2020 Conference on Empirical Methods in Natural
  Language Processing: Findings}, 2020, pp. 4163--4174.

\bibitem{catak2021data}
F.~O. Catak, J.~Ahmed, K.~Sahinbas, and Z.~H. Khand, ``Data augmentation based
  malware detection using convolutional neural networks,'' \emph{PeerJ Computer
  Science}, vol.~7, p. e346, 2021.

\bibitem{millar2021multi}
S.~Millar, N.~McLaughlin, J.~M. del Rincon, and P.~Miller, ``Multi-view deep
  learning for zero-day android malware detection,'' \emph{Journal of
  Information Security and Applications}, vol.~58, p. 102718, 2021.

\bibitem{chilimbi2014project}
T.~Chilimbi, Y.~Suzue, J.~Apacible, and K.~Kalyanaraman, ``Project adam:
  Building an efficient and scalable deep learning training system,'' in
  \emph{11th $\{$USENIX$\}$ Symposium on Operating Systems Design and
  Implementation ($\{$OSDI$\}$ 14)}, 2014, pp. 571--582.

\bibitem{shanmugam2020and}
D.~Shanmugam, D.~Blalock, G.~Balakrishnan, and J.~Guttag, ``When and why
  test-time augmentation works,'' \emph{arXiv preprint arXiv:2011.11156}, 2020.

\bibitem{krizhevsky2012imagenet}
A.~Krizhevsky, I.~Sutskever, and G.~E. Hinton, ``Imagenet classification with
  deep convolutional neural networks,'' \emph{Advances in neural information
  processing systems}, vol.~25, pp. 1097--1105, 2012.

\bibitem{srivastava2014dropout}
N.~Srivastava, G.~Hinton, A.~Krizhevsky, I.~Sutskever, and R.~Salakhutdinov,
  ``Dropout: a simple way to prevent neural networks from overfitting,''
  \emph{The journal of machine learning research}, vol.~15, no.~1, pp.
  1929--1958, 2014.

\bibitem{Bengio:2008}
Y.~Bengio, ``{N}eural net language models,'' \emph{Scholarpedia}, vol.~3,
  no.~1, p. 3881, 2008, revision \#140963.

\bibitem{mikolov2013efficient}
T.~Mikolov, K.~Chen, G.~Corrado, and J.~Dean, ``Efficient estimation of word
  representations in vector space,'' \emph{arXiv preprint arXiv:1301.3781},
  2013.

\bibitem{kingma2014adam}
D.~P. Kingma and J.~Ba, ``Adam: A method for stochastic optimization,''
  \emph{arXiv preprint arXiv:1412.6980}, 2014.

\bibitem{rehurek_lrec}
R.~{\v R}eh{\r u}{\v r}ek and P.~Sojka, ``{Software Framework for Topic
  Modelling with Large Corpora},'' in \emph{{Proceedings of the LREC 2010
  Workshop on New Challenges for NLP Frameworks}}.\hskip 1em plus 0.5em minus
  0.4em\relax ELRA, 2010, pp. 45--50.

\end{thebibliography}

\end{document}